\documentstyle[mncite]{mn2e}
\begin{document}

\title[The cD radio galaxy of Abell 2390]{The radio properties of the cD galaxy of Abell 2390}
\author[Augusto, Edge, and Chandler]{Pedro~Augusto,$^{1,2}$ Alastair~C.~Edge,$^3$
 and Claire~J.~Chandler$^4$\\
$^1$Centro de Astrof\'{\i}sica da Universidade do Porto, Rua do Campo Alegre, 823, 4150 Porto, Portugal\\ 
$^{2}$Universidade da Madeira, Centro de Ci\^encias Matem\'aticas, Caminho da Penteada, 9050 Funchal, Portugal\\
$^3$University of Durham, Dep.\ of Physics, South Road, Durham DH1 3LE, UK\\ 
$^{4}$National Radio Astronomy Observatory, P.O.\ Box O, Socorro, NM 87801,USA} 


\maketitle

\begin{abstract}
We present multi-frequency, multi-epoch radio imaging of the complex
radio source B2151+174 in the core of the cluster, Abell~2390 ($z\simeq0.23$). From
new and literature data we conclude that the FRII-powerful radio
source is the combination of a compact, core-dominated `medium-symmetric object' (MSO)
 with a more extended, steeper spectrum mini-halo.
B2151+174 is unusual in a number of important aspects: i) it is one
of the most compact and flat spectrum sources in a cluster core known; ii)
it shows a complex, compact twin-jet structure in a north-south orientation; iii) the orientation
of the jets is 45\degr misaligned with apparent structure (ionization cones and dust disk) of the host galaxy on larger scales.
Since the twin-jet of the MSO has its northern half with an apparent `twist', it might be that precession of the central supermassive black hole explains this misalignement.
 B2151+174 may be an
example of the early stage (10$^{3-4}$yrs duration) of a `bubble' being blown into the
ICM where the plasma has yet to expand. 
\end{abstract}

\begin{keywords}
galaxies: clusters: individual: Abell 2390 --- galaxies: active --- galaxies: cD.
\end{keywords}
 

\section{Introduction}

The cores of clusters of galaxies host many of the most luminous
radio sources in the local Universe (e.g. Cygnus-A, Hercules-A, Hydra-A).
The influence of such powerful sources of radiation and mechanical
energy on the surrounding intracluster medium is clear in the
core of the Perseus cluster \cite{Bohetal93,Fabetal03}
and may be a significant energy source regulating cooling flows \cite{Soker01,Fabetal02,BrugK02,Ruszet02}.
The majority of this energy is believed to be injected into the
ICM through mechanical work done by radio jets but relatively
few cooling flows have luminous, lobed radio emission and most are
relatively compact and low luminosity. This is as would be 
expected if these radio sources have a duty-cycle of the order
of 10$^8$yr \cite{dlVetal04} but in a large sample of clusters a few
luminous, compact sources should be found that are in the process
of launching jets. 
With a view to investigating the influence of more compact radio
sources on the ICM, we have undertaken an extensive, high resolution
radio imaging campaign of the radio source B2151+174 (\pcite{Broetal98})
that lies in the central galaxy of the rich cluster Abell 2390.
This cluster has a high X-ray luminosity ($L=2.125 \times 10^{45}$ erg~s$^{-1}$  (0.1-2.4 keV) 
--- \pcite{Ebeetal96}) and is known to contain a cooling flow 
($\la 300$ M$_{\odot}$yr$^{-1}$ \cite{Alletal01} from recent Chandra observations). 
The central host galaxy of B2151+174 has extended optical emission lines \cite{LeBetal91},
extended Lyman-$\alpha$ \cite{HutBal00}, strong dust absorption in the optical 
and sub-mm dust continuum emission \cite{Edgetal99}.  B2151+174
is one of the most luminous radio sources in a BCG (FRII 1.4~GHz power of $10^{25.1}$~W/Hz) in the combined BCS/eBCS X-ray flux-limited
sample of 300 clusters \cite{Ebeetal98,Ebeetal00} and samples of radio
sources in cluster cores  \cite{BalBurLok93} but
the radio properties of B2151+174 differentiate it from most canonical
central cluster radio sources. It has an apparently self-absorbed radio
spectrum \cite{Edgetal99} implying that the majority of the radio emission
is from a very compact region.

The kpc-scale radio structure of B2151+174 was unveiled for the 
first time by Augusto et al. (1998), from VLBA  and MERLIN 5 GHz maps, 
who studied it as a candidate compact-symmetric object (CSO). 
In fact, in this paper, we revise the classification and find the source 
to be larger than 1~kpc, making it a medium-sized symmetric object (MSO).
CSOs (sizes  on 1--1000~pc) are radio sources with lobes on either side of a 
compact core which dominates at high frequencies
 (e.g.\ \pcite{Wiletal94}).  MSOs are similar  but larger (1--15~kpc).
CSOs,  short and `twin-sided', are likely young sources ($10^3$--$10^4$ yrs); 
the ages of some have been determined by either kinematical
studies (e.g.\ \pcite{OwsConPol98}) or  
synchrotron emission aging  \cite{Reaetal96a}.



In Section~2 we present the data collection and describe its processing, including the problems
found. In Section~3 we use all the data to make an extensive study of the radio properties of 
B2151+174, from maps, spectra (total and main components), variability (or lack of it), jets 
orientation, size, mini-halo evidence, etc. Finally, in Section~4, we conclude the paper with an overall discussion.
  
Throughout this paper, we assume H$_{0}=75$ km s$^{-1}$ Mpc$^{-1}$, q$_{0}=0.5$ and $\Lambda=0$.



\section{Observations and Data reduction}

 B2151+174 was observed by us 
with several radio interferometer arrays as detailed in Table~\ref{observ}.
All radio data were processed using AIPS (National Radio Astronomy Observatory) 
and DIFMAP (CalTech package) through the standard routines. During the observations 
we used the standard point source and flux calibrators (except for MERLIN 23 GHz as noted below)
as well as fringe finders for the VLBI (VLBA and EVN) observations and a nearby phase-referencing 
calibrator for some of the  observations (see Table~1).

Although we had same epoch MERLIN+EVN 18 cm observations (Table~\ref{observ}), we have 
preferred to keep them separate (MERLIN and EVN) on the context of the detailed analysis done in Section~3.

Augusto et al. (1998) have studied B2151+174 as part of a radio search for gravitational lenses. They  
presented two C-band (5 GHz) maps: one with MERLIN 5 GHz and one with VLBA 5 GHz. 
We have re-reduced the Augusto et al.\ 5~GHz data and obtained consistent results.
We have also re-reduced the VLA-A 8.4~GHz data from the Jodrell-VLA Astrometric Survey (JVAS; \pcite{Broetal98}).

We have produced new MERLIN 23 GHz self-calibrated maps from data mentioned by Augusto et al. (1998) which
had poor phase-reference and self-calibration was not attempted given that the
aim of that paper was a lens search. 
We believe the phase-referencing was unsuccessful because  the map position is actually about $120$ mas 
off the known $\sim$mas-precision position that we use throughout this paper 
(from VLA-A 8.4 GHz data --- \pcite{Broetal98}).
As regards the flux density calibration, unfortunately, the standard 3C286 flux 
calibrator was not observed. Instead, the highly variable point source calibrator 3C273
was used.
Flux monitoring of this source made with MERLIN at 
the time, give or take a few days, gave too high a flux density value for B2151+174 when compared with other K-band data that we have also collected (see below).
This calibration problem affects all Augusto et al.\ MERLIN 23 GHz observations although we
are confident that they have presented trustworthy maps of the five other sources.
 Judging from their overall spectra alone, however, of the five sources in Augusto et al. (1998), two seem to have a too high 
flux density point from MERLIN 23 GHz in the rough proportion that our B2151+174 value 
is high. For the quantitative results of Section~3, we 
recalibrated these MERLIN 23 GHz data using our VLA-DnC 22 GHz measurements (Table~\ref{observ}), by dividing the initial flux densities by 3.05.

We have decided to either use maximum resolution (uniform weighting of baselines) 
or maximum sensitivity (natural weighting), depending on the best option for the detailed 
component analysis that followed. These were as follows: i) uniform weighting --- MERLIN 1.7 GHz, 
VLA-A 8.4 GHz, VLA-A 43 GHz; ii) natural weighting ---  EVN 1.7 GHz, MERLIN 5 GHz, MERLIN 23 GHz,  VLBA 5 GHz.
 In all maps produced we have used the three-sigma level for the lowest contour (where sigma 
is the r.m.s on the map shown, in each case, at the relevant caption). See Table~\ref{details} which contains all map details.


\begin{table*}
\caption{The radio observations of B2151+174, RA (2000) = 21$^h$ 53$^m$ 36\fs8267, 
Dec (2000) = +17\degr 41\arcmin 43\farcs726, ordered chronologically.  All are continuum observations. 
The observing time is given `on-source' since for phase-referenced observations the actual 
total run was longer. The resolutions presented here are just approximate. When others used the same data, we give the relevant reference. $^*$The MERLIN+EVN used the following telescopes: LoDeCbKnDaMkTa + EfJbCbMcNtOnWbTr.}
\begin{tabular}{llccccc}
\hline
\multicolumn{1}{c}{Telescope}&\multicolumn{1}{c}{Observing}&  Integration &Phase-refd.\  &Frequency      & Resolution & Reference \\
         &\multicolumn{1}{c}{date}     &time (on-source)  & observations?   & (GHz) &(arcsec) &   \\
\hline
VLA-A &1992 Oct 18 & 2 min & yes &8.4 &0.2 & \scite{Broetal98} \\
VLA-D & 1993 Sep 7 & 10 sec  & yes & 1.4 & 45 & \scite{Conetal98} \\
MERLIN&1995 Jun 4& 80 min & yes &5.0 &0.05& \scite{papI} \\ 
MERLIN & 1996 May 15 & 4 hrs & ``yes'' (failed) & 23 & 0.015 & \\
VLBA+Y &1996 Jul 25&1 hr & no &5.0 &0.003 & \scite{papI} \\
MERLIN+EVN$^*$&1997 May 30 &1.5 hrs & yes &1.7 &0.02 & \\ 
VLA-DnC &1997 Oct 22 & 16 min & yes & 1.4 & 226.0  & \scite{Edgetal99} \\ 
VLA-DnC &1997 Oct 22 & 16 min  &  yes & 4.85 & 210.8 &\scite{Edgetal99} \\ 
VLA-DnC &1997 Oct 22 & 6 min  & yes & 8.4 & 139.4 &\scite{Edgetal99} \\ 
VLA-DnC &1997 Oct 22 & 19 min & yes & 15 & 101.1 &\scite{Edgetal99} \\ 
VLA-DnC &1997 Oct 22 & 26 min & yes & 22 & 73.2  &\scite{Edgetal99} \\ 
VLA-DnC &1997 Oct 22 & 37 min & yes & 43 &  41.7 &\scite{Edgetal99} \\ 
VLA-A &2002 Feb 12 & 2 min & no &43 &0.06 & \\ 
VLA-B & 2004 May 9 & $\approx$1hr & yes & 0.074 & 80 & lwa.nrl.navy.mil/VLSS \\
\hline 
\end{tabular} 
\label{observ}
\end{table*} 

\begin{table*}
\caption{The details for each of the maps of B2151+174 in Figures~1 and~2. The contours are drawn in powers of two from the first one.
Note that the peak flux density always decreases with smaller beam size. The asterisk marks recalibrated flux density values (see Section~2).}
\begin{tabular}{ccccccc}
\hline \hline
Instrument & Freq. & Fig. & beam  & peak & 1st contour (3$\sigma$) & Total flux \\
 & (GHz) & & ($\times0\farcs001$)  & (mJy/beam) & (\%) & density (mJy) \\
\hline
MERLIN & 1.7 & 1c & $90\times90$  & 141 & 0.7 & 219 $\pm$20 \\
EVN & 1.7 & 2c & $20\times20$  & 62 & 2 & 171 $\pm$26 \\
MERLIN & 5.0 & 1a & $70\times70$  & 168 & 0.33 & 199 $\pm$20 \\
VLBA & 5.0 & 2a & $2.5\times2.5$  & 115 & 0.5 & 157 $\pm$14 \\
VLBA & 5.0 & 2b & $10\times10$  & 128 & 0.5 & 157 $\pm$14 \\
VLA-A & 8.4 & 1d & $90\times90$  & 116 & 0.7 & 134 $\pm$23 \\
MERLIN & 23 & 2d & $15\times15$  & 45$^*$ & 2.5 & 73$^*$ $\pm$24 \\
VLA-A & 43 & 1b & $40\times40$  & 45 & 8.5 & 61 $\pm$15 \\
\hline
\end{tabular} 
\label{details}
\end{table*}

\section{Radio Properties of the cD}

\begin{figure*} 
\setlength{\unitlength}{1cm}
\begin{picture}(16,22)
  \put(-3.5,9.5){\includegraphics{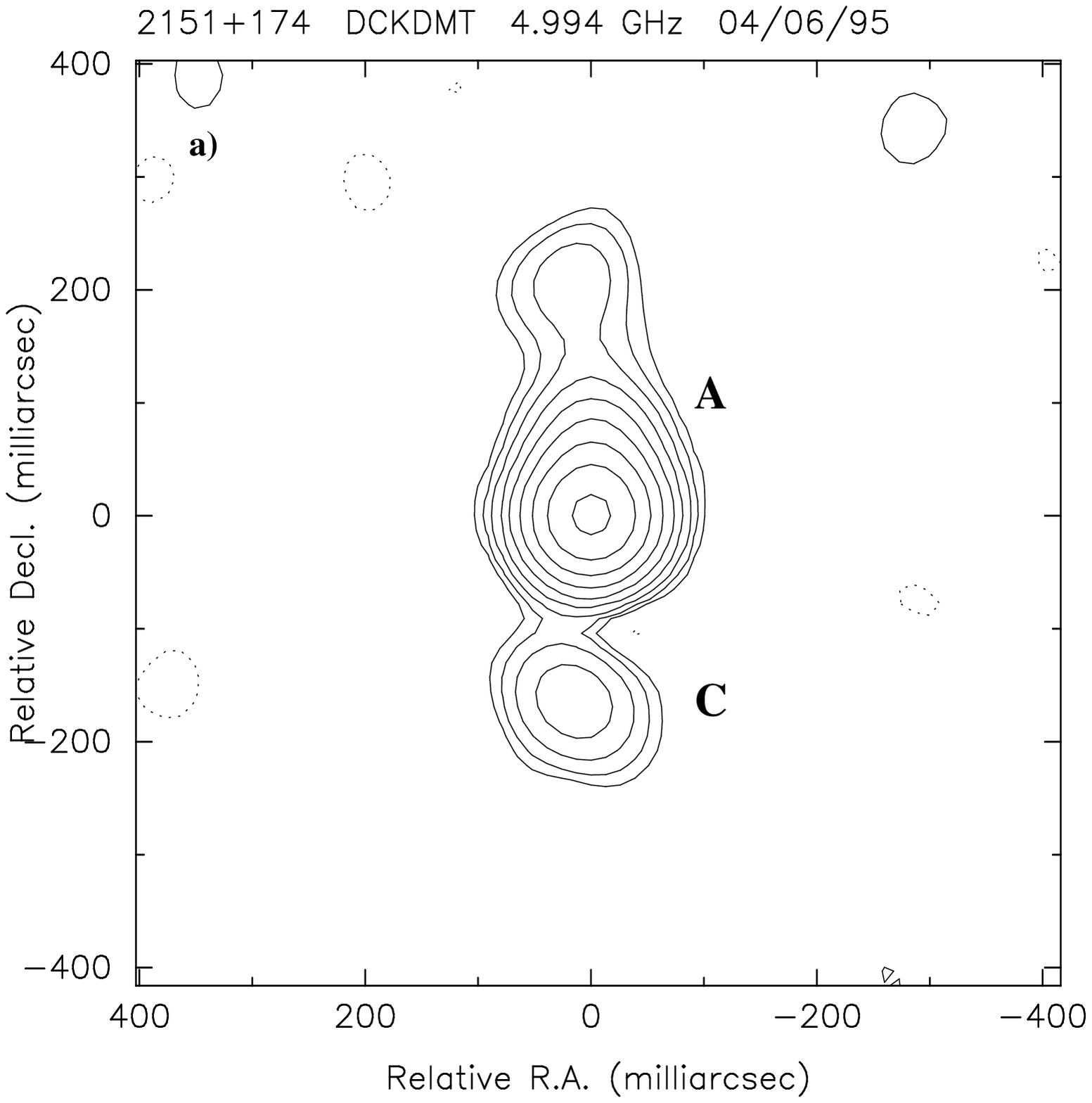}}
  \put(5.3,9.8){\includegraphics{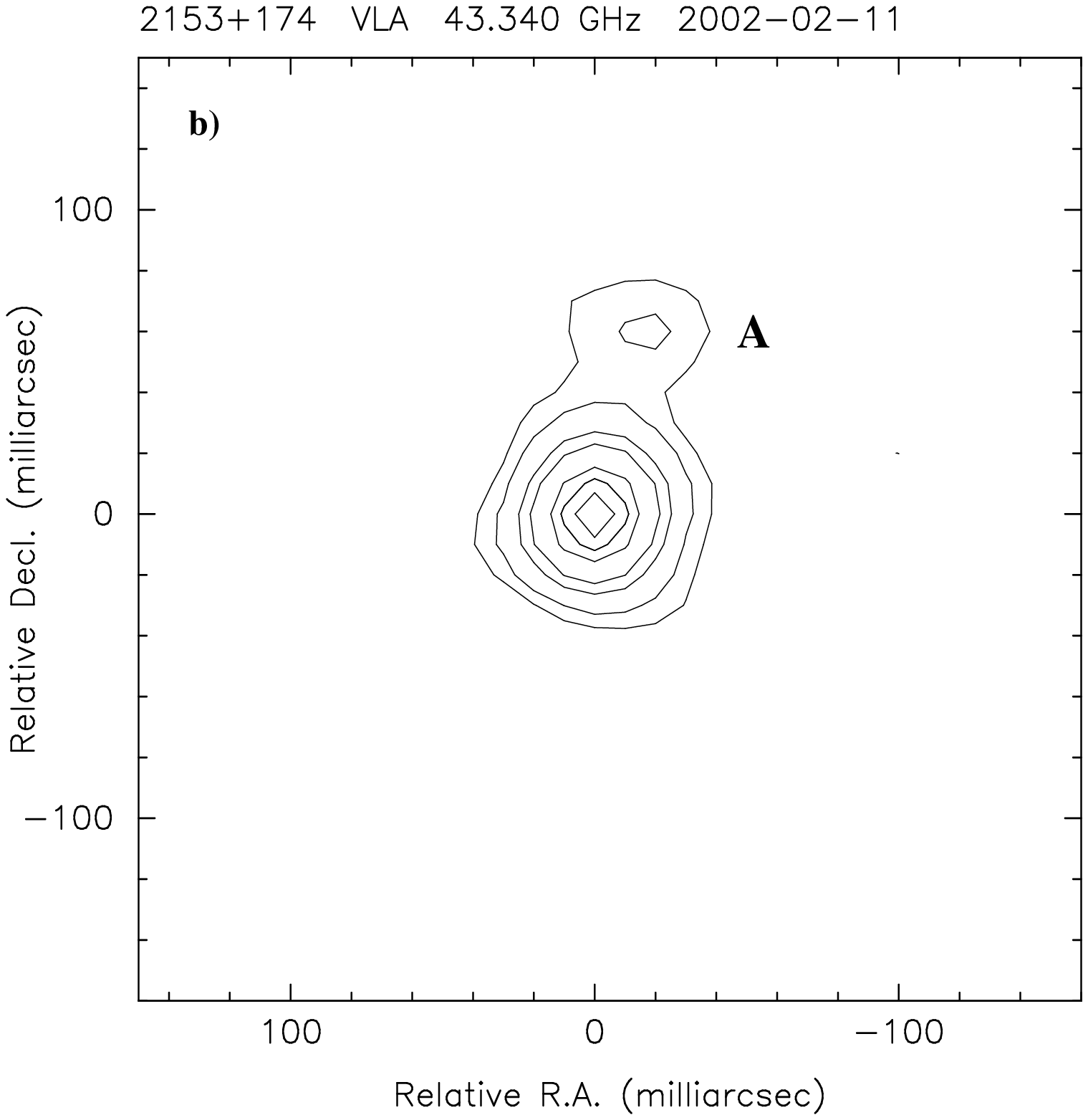}}
  \put(-3.5,-1){\includegraphics{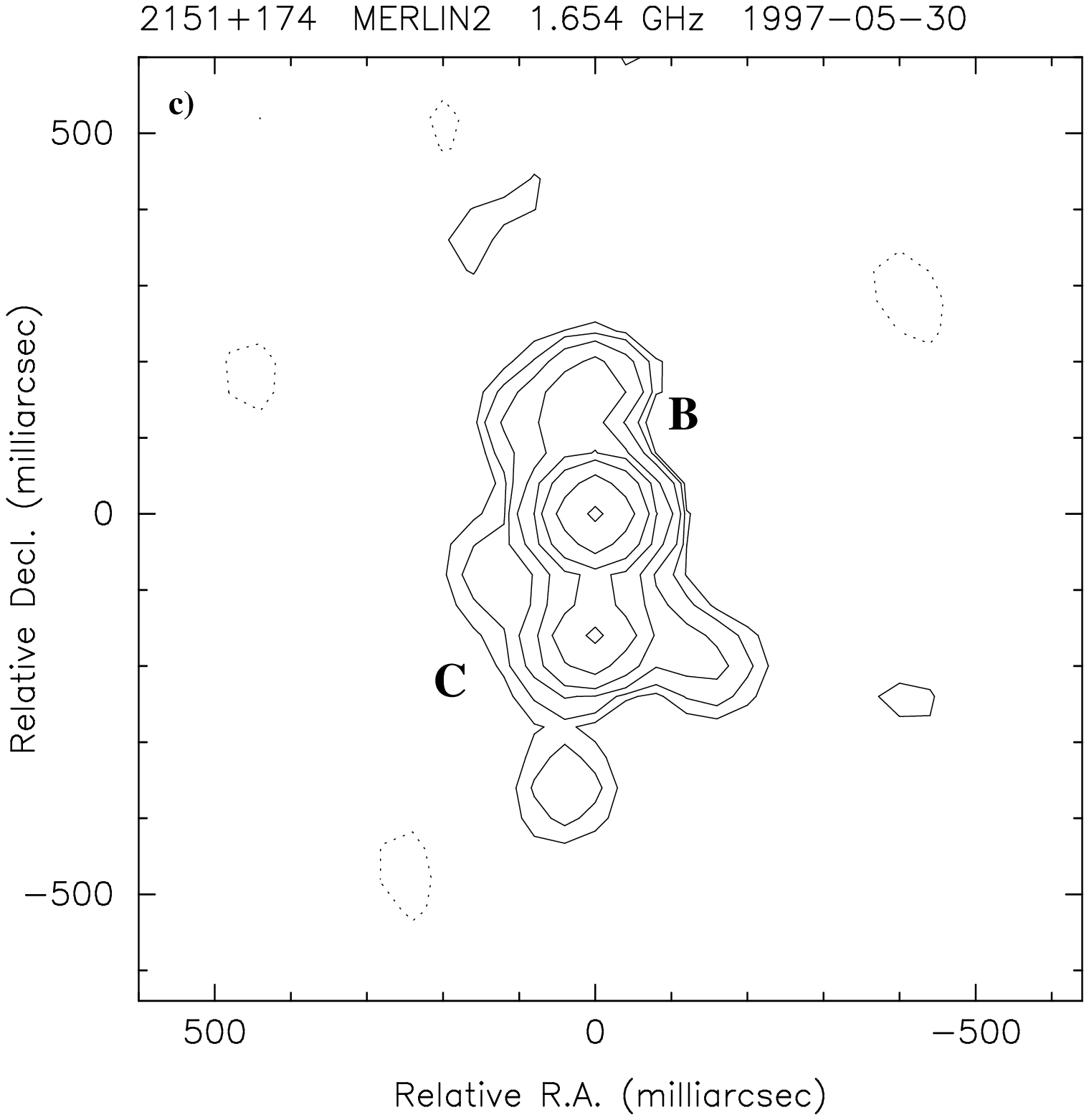}}
  \put(5.3,-1.2){\includegraphics{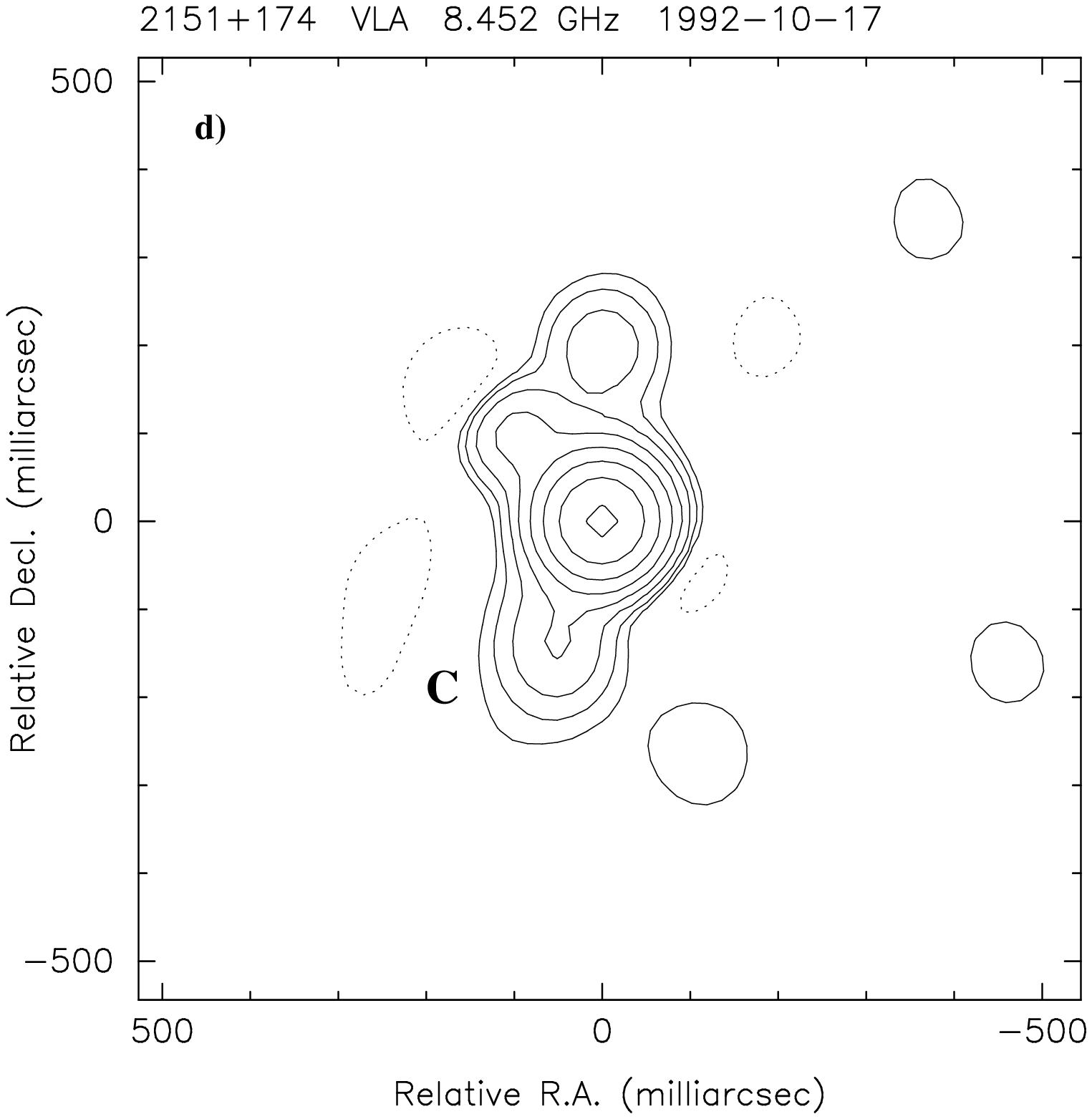}}
\end{picture}
\caption{ The maps produced with all B2151+174 data containing labeled non-nuclear components as in Table~3:
a) MERLIN 5 GHz map, as in Augusto et al. (1998) but convolved with a 
70 mas circular beam; b)  VLA-A 43 GHz map where the data were convolved with a 40 mas circular beam; c) 
MERLIN 1.7~GHz map; the  data were convolved with a 90 mas 
circular beam (super-resolved by a factor of 1.3--2, depending on direction); d) VLA-A 8.4 GHz map with the data convolved 
with a 90 mas circular beam (super-resolved by a factor of 1.9--2.3, depending on direction). 
See Table~2 for the details of each map. 
} 
\label{labeled_comps}
\end{figure*}

\begin{figure*} 
\setlength{\unitlength}{1cm}
\begin{picture}(16,22)
  \put(-3.5,9.8){\includegraphics{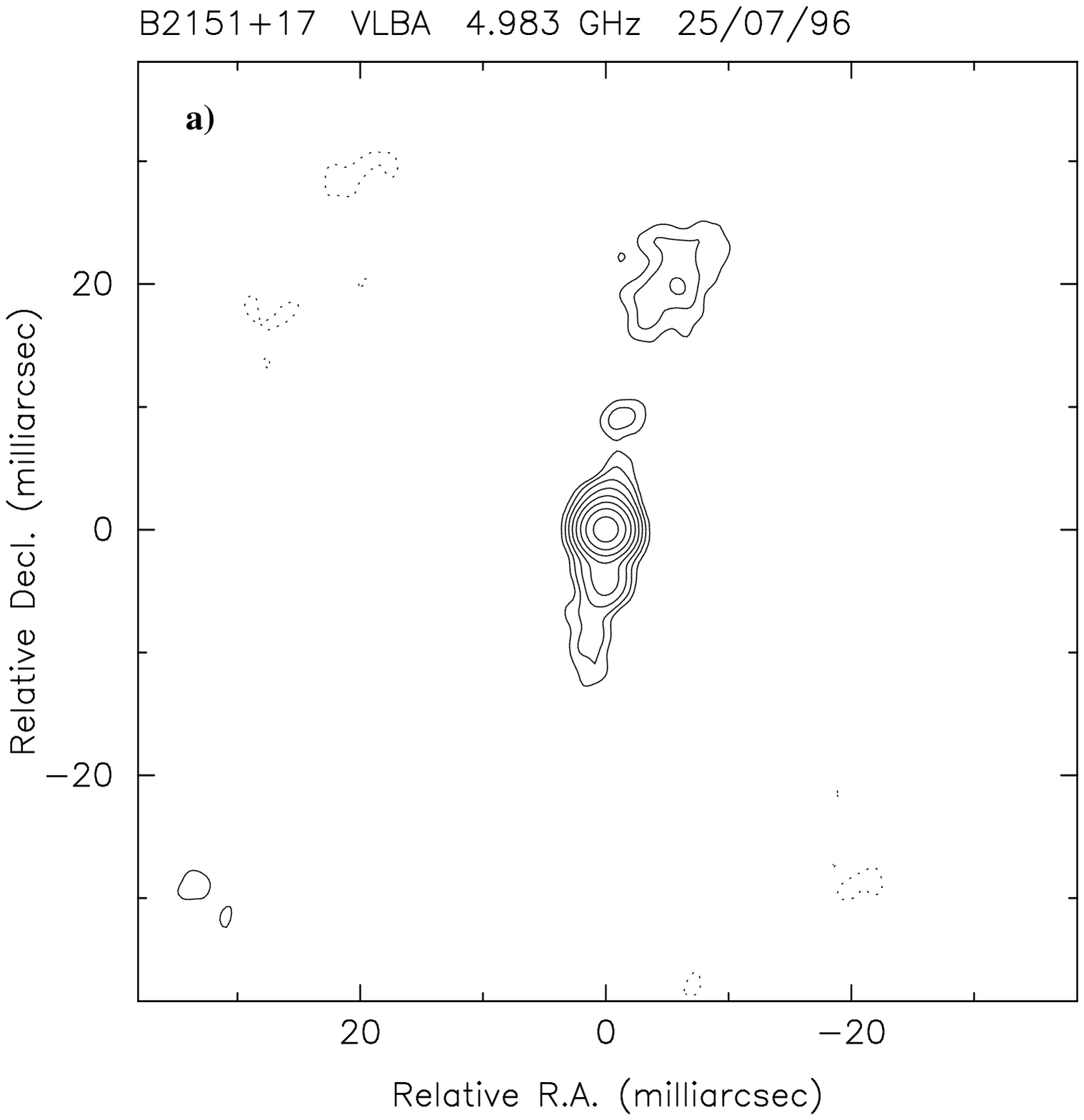}}
  \put(5.3,9.8){\includegraphics{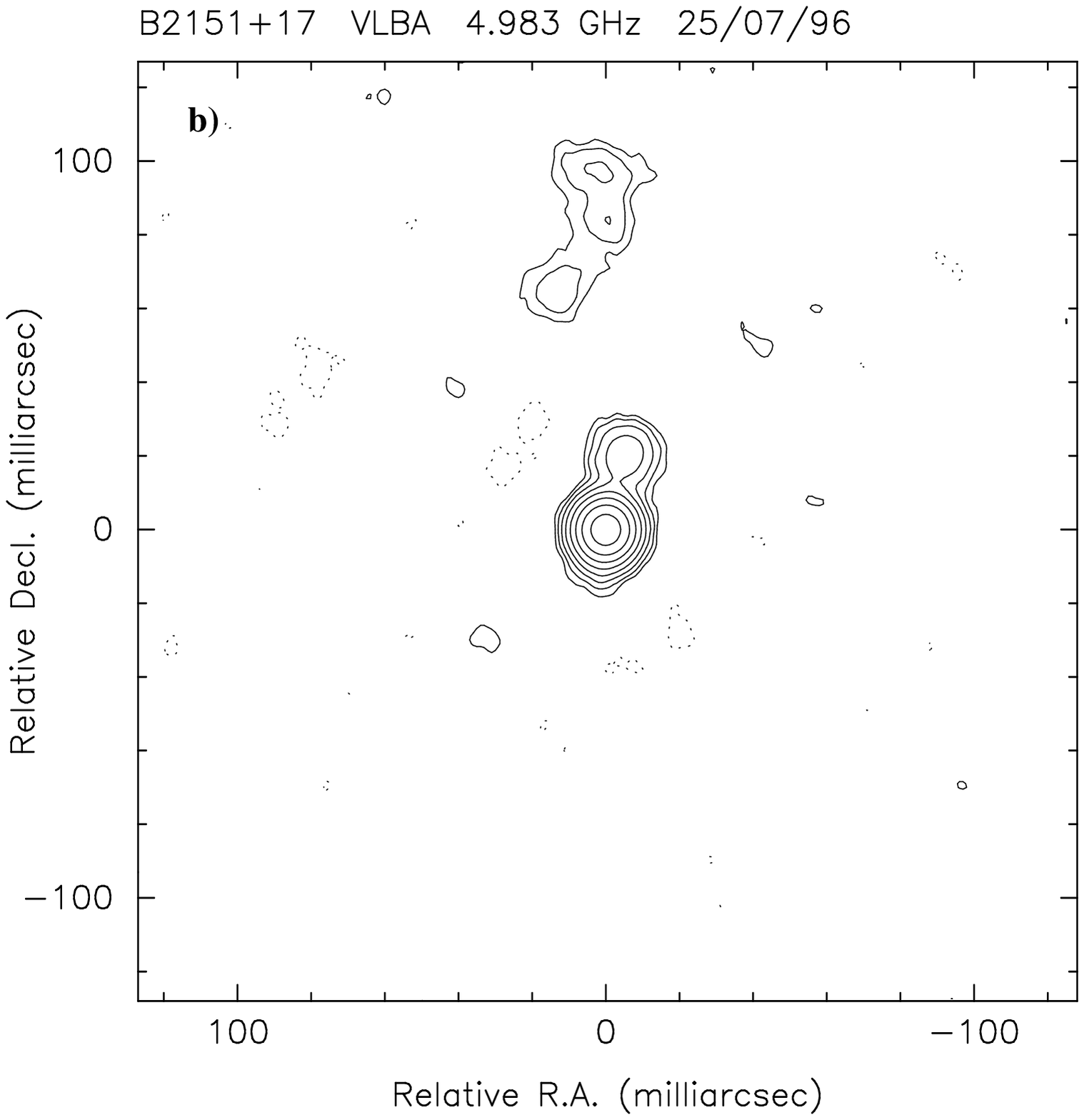}}
  \put(-3.5,-1){\includegraphics{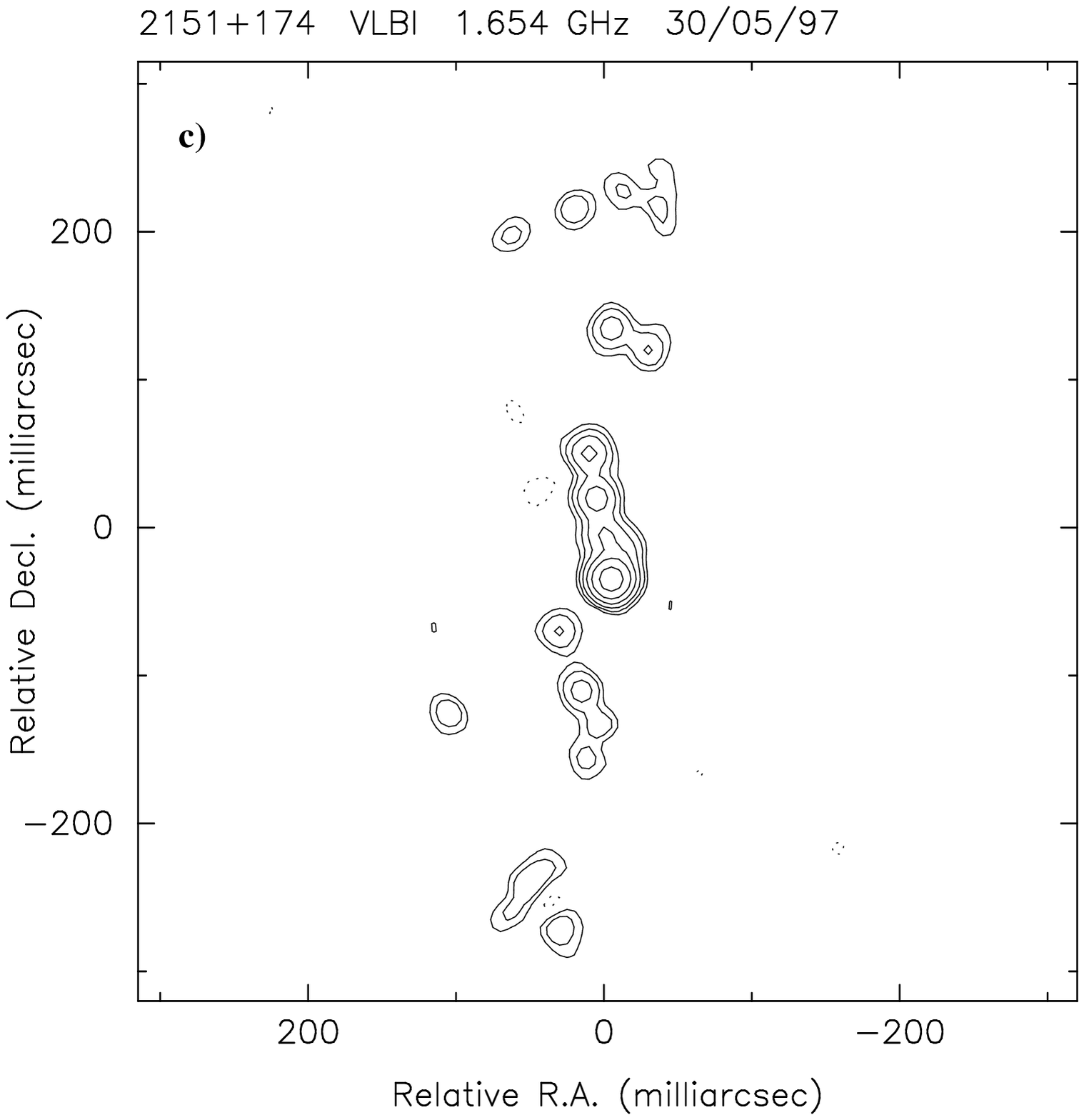}}
  \put(5.3,-1){\includegraphics{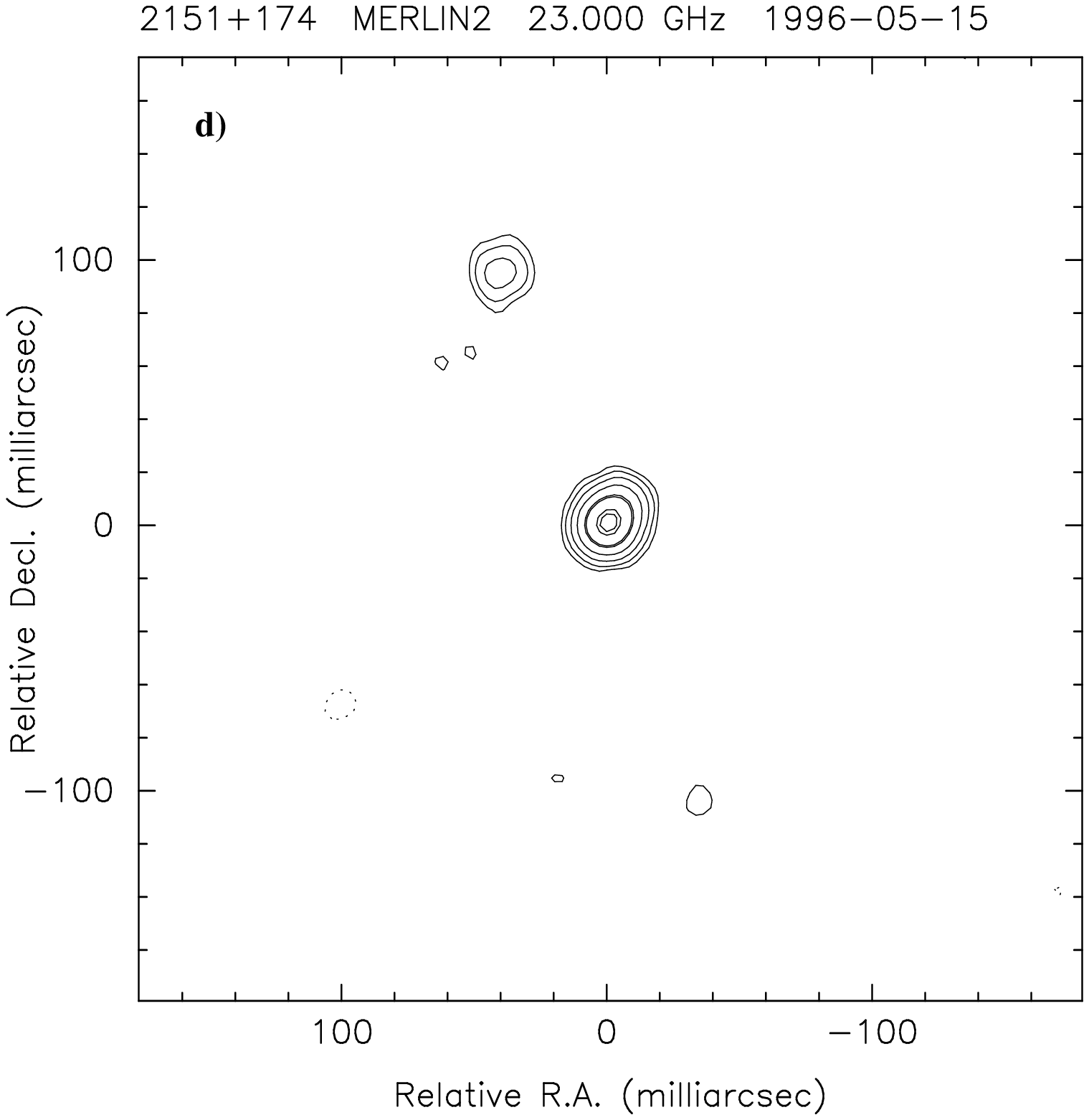}}
\end{picture}
\caption{
The maps produced with all B2151+174 data with only nuclear components  labeled in Table~3:
a) VLBA 5 GHz map, with the data convolved with a 2.5 mas circular beam --- only the central nuclear region is shown; 
b) VLBA 5 GHz map produced by convolving the data with a 10 mas circular beam; the central region (shown in a)) is at the centre; c) EVN 1.7~GHz map with a 20 mas circular beam; d) MERLIN 23 GHz map with a 15 mas circular beam.
See Table~2 for the details of each map. 
} 
\label{no_labeled_comps}
\end{figure*}

We have model fitted all maps in DIFMAP and present the resulting models in Table~\ref{Tabmod}. 
With the objective of a confident description of the structure and spectra of the B2151+174 
components, we were careful to use several ways of checking that the models were sensible: 
i) using DIFMAP to inspect how well the model was fitting the data both in visibility and residuals; 
ii) plotting the corresponding maps and comparing them with the ones obtained by the CLEANing process 
in DIFMAP (Figures~~\ref{labeled_comps} and~\ref{no_labeled_comps}); iii) comparing the r.m.s. of the CLEANed maps with the r.m.s of the 
maps obtained with the corresponding models: the models all have an r.m.s which is within 50\% of the one of the CLEANed maps.
For models with $\chi^2>1$ (VLBA 5~GHz and EVN 1.7~GHz) only the strongest (nuclear) component was used in this Section.
Furthermore, in the following subsections we have not considered
 the components that had: 
i) a surface brightness less than 20 times the r.m.s. surface brightness (for which we used the beam sizes); 
ii)  a fitted axial ratio of zero, {\em unless} they were a nuclear component or the semimajor axis of the 
component was smaller than the semimajor axis of the respective beam. 
Apart from the nucleus (clearly identified in {\em all} datasets) we only used three components (labeled A, B and C) --- Table~3. When
the fitted ellipses intersected, we cross-identified the same component in different datasets (see Figure~1).

\begin{table}
\caption{The models fitted to all radio data with maps in Figures~1 and~2. For each gaussian component, 
$S$ is the flux density ($^*$ indicates recalibration --- see Section~2), $r$ and $\theta$ describe the separation vector from the origin, $a$ is the 
major axis, $b/a$ is the axial ratio and $\Phi$ is the position angle of elongation. Finally, $\chi^2$ 
is the error in the fitting, well described in Polatidis et al. (1995): if less than one the model is fine. The components on which we are  most confident for spectra and 
morphology analysis are labeled with the corresponding letter (in the final column) as in the maps.}
\begin{center}
\begin{tabular}{rrrrrrr} \hline
 \multicolumn{1}{c}{$S$} & \multicolumn{1}{c}{$r$} & \multicolumn{1}{c}{$\theta$} & \multicolumn{1}{c}{$a$} & \multicolumn{1}{c}{$b/a$} & \multicolumn{1}{c}{$\Phi$} & \multicolumn{1}{c}{$\chi^2$}  \\
  \multicolumn{1}{c}{(mJy)} & \multicolumn{1}{c}{(mas)} & \multicolumn{1}{c}{($^{\circ}$)} & \multicolumn{1}{c}{(mas)} & & \multicolumn{1}{c}{($^{\circ}$)} & comp. \\ \hline \hline
\multicolumn{6}{c}{\bf MERLIN 1.7 GHz} & 0.73 \\ \hline
138 &    0.0&     0.0&      65.1& 0.00&    16.6& N--large \\
 35&     140.9&    174.3&     116.7&    0.64&   $-$5.5& C \\
 23&     155.2&    5.8&    127.3&    0.82&   $-$44.0& B\\
 15&     187.2&    $-$152.2&     298.8&    0.67&    70.9& \\
7&     285.1&     176.4&     310.6&     0.00&   $-$22.7& \\ \hline
\multicolumn{6}{c}{\bf EVN 1.7 GHz} & 2.26 \\ \hline
63&     0.0&      0.0&      7.3&    0.13&    56.2& N--small \\
40&     44.8&     6.4&     60.2&     0.25&    17.3& \\
 20&     73.2&     160.3&     78.7&   0.10&    20.0& \\
 18&    171.3&     $-$0.6&     26.7& 0.67&   7.3& \\
 12&     196.8&     165.2&     116.3& 0.00&    $-$27.6& \\
 10&    237.9&   $-$5.8&    114.6&    0.00&   $-$22.7& \\ \hline
\multicolumn{6}{c}{\bf MERLIN 5.0 GHz} & 0.72 \\ \hline
 164& 0.0&      0.0&      16.3& 0.00&   0.4& N--large \\
 25&     64.2&    1.1&     52.7& 0.00&   0.33& A\\   
7&    141.2&     176.6&     60.7&   0.17&   $-$21.0&C \\
4&     269.7&   $-$0.4&     192.9&     0.00&   $-$6.3& \\  \\ \hline
\multicolumn{6}{c}{\bf VLBA 5.0 GHz} & 2.33 \\ \hline
121&    0.0&     0.0&     0.8&    0.32&   $-$3.9& N--small \\
 30&     80.0&    1.4&     50.0&    0.1&   $-$26.0& \\
8&     4.7&     179.0&     2.2&    0.38&   $-$10.6& \\ \hline
\multicolumn{6}{c}{\bf VLA-A 8.4 GHz} & 0.00058 \\ \hline
115&    0.0&      0.0&      49.6&     0.00&  $-$9.2& N--large \\
9&     78.3&     42.5&     182.0& 0.00&    11.5& \\
8&     153.0&     167.5&    200.9&   0.32&    10.1&C \\
5&    160.2&     3.6&     251.1& 0.00&   $-$8.4& \\ \hline
\multicolumn{6}{c}{\bf MERLIN 23 GHz} & 0.33 \\ \hline
48$^*$ &       0.0 & 0.0 &    8.0&        0.57&        16.3& N--small \\
15$^*$&     104.8&   22.9&   11.5&  0.66&        $-$25.8& \\
8$^*$&     9.7&         $-$32.4&        11.9&         0.00& $-$12.0& \\ \hline
\multicolumn{6}{c}{\bf VLA-A 43 GHz} &  0.36 \\ \hline
42&     0.0&      0.0&      15.3&    0.15&    11.2& N--small\\
6&     64.7&    $-$21.8&     83.8&    0.00&  $-$77.9& A\\ \hline
\end{tabular}
\end{center}
\label{Tabmod}
\end{table}

\subsection{Morphology}

 The closest component to the nucleus is A, seen in 
both MERLIN~5~GHz and VLA-A 43~GHz data --- see  Figures~1a,1b and Table~3. Making the average between both models, this component is $\sim64$~mas ($\sim0.20$~kpc) from the 
nucleus\footnote{In all that follows we assume the radio structure to be perpendicular to our line-of-sight, 
i.e., parallel to the plane of the sky, so all quoted distances are linear {\em projected} distances. 
} at p.a. $\simeq-11\degr$ and compact (smaller than the formal resolution of both datasets and with 0.00 fitted axial ratios). 
Further away and on the east side of north resides component B, only identified with MERLIN~1.7~GHz (Figure~1c; see also 
 Table~3). This one is $\sim155$~mas ($\sim0.48$~kpc) from the nucleus at p.a. $\simeq+6\degr$ and is clearly resolved 
(an almost circular `blob' $\sim127$~mas ($\sim0.40$~kpc) in diameter). Finally, on the other side of the nucleus and at a similar 
distance (but not quite opposed) lies component C, seen with MERLIN~5~GHz, MERLIN~1.7~GHz, and  VLA-A~8.4~GHz (Figures~1a,1c,1d) $\sim145$~mas ($\sim0.45$~kpc) away from the nucleus 
at p.a. $\simeq+173\degr$ (again taking the average values). This is also a resolved `blob', although the different resolutions 
cannot clearly reveal its size; it is elliptical in shape (somewhere on $\sim60$--200~mas ($\sim0.2$--0.6~kpc) with a 0.2--0.6 axial ratio).


As regards the three highest resolution maps (EVN~1.7~GHz, VLBA~5.0~GHz, and MERLIN~23~GHz --- Figure~2) the structures seen
 support the overall double-and-opposite-jet structure seen with the labeled components: the MERLIN~23~GHz map hints at a bending jet, starting off at the NW and bending afterwards to NE; the EVN~1.7~GHz map is consistent with this, but only shows larger scale structure on both sides of the nucleus: towards SE and towards NE --- most components are too resolved; the VLBA~5.0~GHz map, with a finer resolution than any of the previous ones, similarly to the EVN data hints at a bending northern jet, starting off at the NW and bending through north: from the (smaller) distances of the components seen, the whole picture is consistent with both the MERLIN~23~GHz and EVN~1.7~GHz maps. Finally, the VLBA~5.0~GHz map also shows, very close to the nucleus, a SE pointing jet.

 B2151+174 can be classified as a small bright core medium symmetric object 
(MSO), given its triple component structure which is dominated by a bright core (e.g. \pcite{papI,Augetal99}).
Using the same `5~GHz 6--$\sigma$' size definition for CSO/MSOs of Augusto et al. (2005), the MERLIN 5.0~GHz map implies an angular size of 0.49\arcsec for B2151+174 or 1.5~kpc at its
  z=0.2302  \cite{LeBetal91,Yeeetal96}. In fact, at 1.7~GHz with MERLIN the size is almost twice, using the same convention (2.7~kpc).

\subsection{The spectra of B2151+174 and its main components}

\begin{figure*} 
\setlength{\unitlength}{1cm}
\begin{picture}(16,11)
 \put(0,12.5){\includegraphics{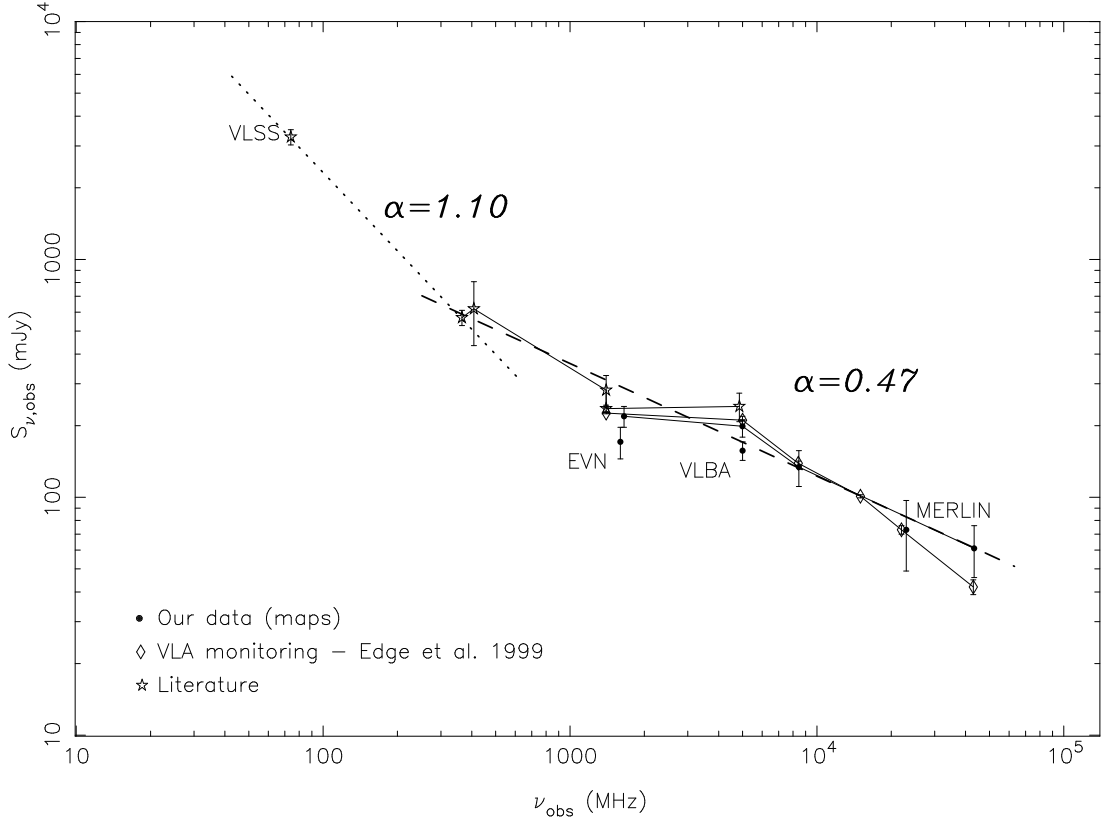}}
 \end{picture}
\caption{The total spectra of B2151+174. In this plot we show all available data points 
from three different sources: first, from our data (\protect Table~\ref{observ} and 
Section~3.1); second, from the VLA monitoring we conducted with results published in 
Edge et al. (1999); finally, from the literature, whose different references 
are given in the text. There are some other notes worthy of mention. First, note the three points that are not connected with lines and 
correspond to high resolution observations (EVN, VLBA, and MERLIN 23~GHz) which missed 
extended structure seen with MERLIN at 1.7 and 5~GHz
(black circles with  lines connected). 
Our VLA-A 8.4 GHz measurement ($134 \pm 23$ mJy) is entirely consistent with the 
one of Browne et al. (1998), 129 mJy, and with Edge et al. (1999)
VLA-DnC monitoring. Except for the three noted high resolution points and the VLSS one, we have used all the 
remaining to fit the dashed line shown, getting $\alpha\simeq0.47$ ($r^2=0.838$), S$_{\nu} \propto \alpha^{-\nu}$.
Finally, in a dotted line, we also present the two-point $\alpha_{74}^{365} \simeq 1.10$, which is consistent with a mini-halo in Abell 2390 (see Section~3.6).
}
\label{total_spec}
\end{figure*}

We have measured the total flux density of B2151+174 for all continuum data presented in 
Table~\ref{observ} 
and show 
the thus compiled spectra in Figure~\ref{total_spec}. In this same Figure we also 
include, from the literature, the following points (in frequency order): 
74~MHz (VLSS),365 MHz  \cite{Douetal96}, 408 MHz \cite{Dix70}, 1400 MHz \cite{WhiBec92,Conetal98}, 
and 4850 MHz \cite{GreCon91}.

The overall shape of the spectrum is typical of a two-component source, containing a compact portion (dominant at high-frequencies and responsible for the flat $\alpha_{0.365}^{43}\simeq 0.47$) and a very extended one (dominant at low-frequencies and responsible for the steep $\alpha_{74}^{365}\simeq 1.10$). The first is the MSO part, described in the detail in the previous Section, while the latter is the mini-halo, described in Section~3.6.
The spectra, again, becomes steeper on the way to high-$\nu$ radio  and sub-mm wavelengths
where $\alpha \sim 0.7$--0.9 \cite{Edgetal99}.


To get spectral information on the several components of B2151+174, we must decide first 
on how to superpose the different frequency maps. In an ideal situation, one would like 
to have all maps properly phase-referenced and also same epoch. This is not the case 
since the data are both sparse in epochs (a rough 10-year span from the earliest through 
the latest data) and also not all were phase-referenced to a calibrator (Table~\ref{observ}) 
and for some of the ones which were it is not clear that this has worked properly. 
Hence, since this is a source that is core-dominated at all frequencies observed (except the lowest --- 74~MHz), 
we assume that it is safe to take the brightest component on all maps as the central nucleus.

Proceeding from this assumption we can now `superpose' all maps (in fact, the component 
positions of Table~\ref{Tabmod}) to get spectral information for three components:
nucleus (N), A and C.
 We have preferred not to do this in the standard way of producing spectral index maps 
for two reasons. First, we have too many frequencies to play with and spectral index 
maps are (usually) made only between two frequencies. This would imply too many spectral 
index maps to interpret for such a relatively simple (and weak) radio source. 
Second, related to this weakness of the structure, none of our datasets was 
actually a full run of observations. Most were short shapshots with poor $u$--$v$ coverage 
(the poor quality of the data is reflected in the $\sim40$\% of the components in 
Table~\ref{Tabmod} which are not labeled). 

\begin{figure*}
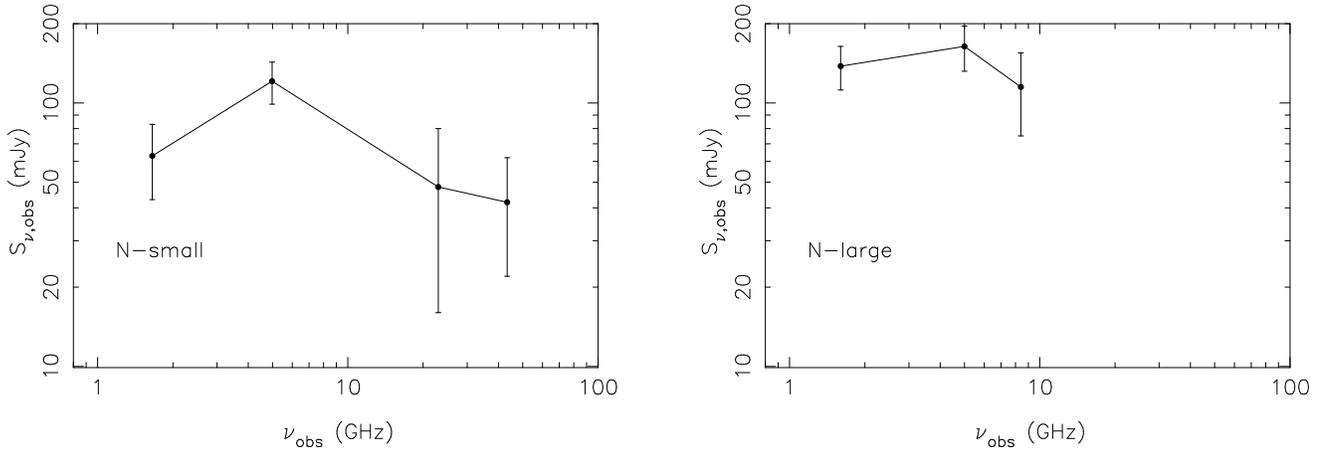
 
\setlength{\unitlength}{1cm}
\begin{picture}(8,6.5)
 \put(-5.4,7.5){\includegraphics{Fig4a.eps}}
  \put(3.8,7.5){\includegraphics{Fig4b.eps}}
\end{picture}
\caption{The spectrum of the nucleus of B2151+174 as model fitted using EVN 1.7 GHz, 
VLBA 5 GHz, MERLIN 23 GHz (recalibrated --- see Section~2) and VLA-A 43 GHz data (N--small) or using MERLIN 1.7 GHz, 
MERLIN 5 GHz and VLA-A 8.4 GHz data (N--large).}
\label{nucl_spec}
\end{figure*}

Since the nuclear component in model fitting had its flux density measured 
by seven instruments with a range  of resolution spanning nearly two orders of magnitude 
(translating into the very different physical sizes), we found it better to split it into 
the `compact nucleus' (N-small), using EVN, VLBA,  
MERLIN 23 GHz (recalibrated --- see Section~2)
and VLA-A 43 GHz data and the `extended nucleus' (N-large), 
using the remaining data. In Figure~\ref{nucl_spec} we present the spectra of both approaches.
The characteristic Giga-Hertz peaked spectrum for N-small leaves no room for doubt that this is a typical AGN-like nucleus.

For component A, we actually have two data points (MERLIN 5 GHz and VLA-A 43 GHz) which give (S$_{\nu} \propto \nu^{-\alpha}$):
$\alpha_{5.0}^{43}$ (A) =0.66$\pm0.2$ (c.f. $\alpha_{5.0}^{43} $ (N) =0.63$\pm0.2$). 
Finally, component C has three-frequency data points (MERLIN 1.7~GHz and 5.0~GHz and VLA-A 8.4~GHz), but the highest frequency 
points are both weak ($<10$ mJy) and similar. Hence, we do not plot its spectra 
but rather get $\alpha_{1.7}^{5.0} $ (C) =1.46$\pm0.2$. 
Thus, it seems that the south component is dominated by a lobe-like steep-spectrum structure (C) while the north shows a compact `knot' (A) on the way to another probable lobe-like structure (B) that we cannot confirm for lack of data. All in all, it is very likely that the north jet of B2151+174 is slightly pointing towards us.

\subsection{Variability}

Our data support the conclusion of Edge et al. (1999) that there has been no significant
variability in  B2151+174 in the past 15 years. Our highest resolution mapping
also shows no significant displacement of the radio components in position (as
would be expected for unbeamed jets viewed close to perpendicular to the line of sight).
It is quite clear (Figure~\ref{nucl_spec})
that 5 GHz is the ideal frequency to probe the compact nuclear component
to test for variability in the future.

\subsection{The age of the radio source and cooling flow timing}

Assuming that the labeled component furthest from the nucleus is travelling at close to the
speed of light (as would be expected from other powerful radio sources), then we
can place a crude lower limit on the timescale of the recent activity. Taking
components B and C (both 0.15$''$ or 0.45~kpc offset from the nucleus) as the
most distant, then this limit is $\simeq1.5\times10^3$~yrs. If we take the estimated full size of B2151+174 (see end of Section~3.1) then this limit increases to $\simeq4.5\times10^3$~yrs.

To assess the validity of this limit we need to determine the synchrotron radiation
timescale ($t_{sync}$). To do this we must first
get an estimate for the magnetic field/energy density (B$_{me}$/u$_{me}$) for each component assuming minimum energy (near equipartition 
of energy between relativistic particles and magnetic field), the same energy in heavy particles 
and electrons, a filling factor of {\it one} for the emitting electrons, cylindrical symmetry and 
an alignment of the magnetic field such that $\sin \phi=1$ \cite{Mil80}, i.e., the magnetic field is in the plane of the sky:

\[
\left\{ 
\begin{array}{l}
B_{me}=0.569  \left[2 \frac{ \textstyle (1+z)^{3+\alpha}}{\textstyle \theta_x \theta_y \, s} 
\frac{\textstyle S_{\nu_0}}{\textstyle \nu_0^{-\alpha}} \frac{ \textstyle \nu_{hi}^{0.5-\alpha} - \nu_{lo}^{0.5-\alpha} }{\textstyle 0.5-\alpha}
\right]^{2/7} \: \: \mu {\rm G} \\
u_{me}=9.28 \times 10^{-14} B_{me}^2 \: erg/cm^3 \: \: {\rm (B_{me} \: in \: \mu G)}
\end{array}
\right.
\]
with $\theta_x,\theta_y$ the dimensions (in arcsec) of each component (model fitted --- 
Table~\ref{Tabmod}), $s=min\{ \theta_x,\theta_y \}$ (kpc) the path length through the 
source in the line-of-sight, $z$ the redshift of the source (0.2302 in this case) and 
the remaining parameters refer to the spectra of the components, many of which had to be guessed; 
$S_{\nu}$ is in Jy and the $\nu$'s in GHz --- more details in the caption of Table~\ref{MagTime} where we present the results.
For the $B_{me}$ calculation, components with fitted axial ratios (in the respective dataset) 
of $0.00$  had this changed to $0.01$ (calculation of $\theta_y$ and $s$).
The $t_{sync}$ calculation was simplified from the formulae used \cite{Muretal99}, ignoring expansion effects
(B$_{CMB}$ is the magnetic field equivalent to the CMB and equals $3.25(1+z)^2$ in $\mu$G):

\begin{table*}
\caption{The physical parameters of the four labeled components (column {\bf 1}) of \protect Table~\ref{Tabmod}
with spectral information in Section~3.2, including the nuclear component N split into two. 
We have used the most appropriate datasets (column {\bf 2}), given the flux density/frequency 
data points (columns {\bf 9--10}) available for each component spectra: the spectral indices 
(column {\bf 8}) were calculated between the shown frequencies, the best approximation possible 
for the formal calculation between $\nu_{hi}$ and	$\nu_{lo}$ (columns {\bf 6--7}) . 
These are, respectively, the two cutoff frequencies (which eventually might be) seen in the 
spectra of each component, the former one most crucial,  indicative of spectral ageing and 
giving a more accurate calculation of the sychrotron age of the emitting electrons 
(column {\bf 13}). For the calculation of this $t_{sync}$ we have guessed $\nu_{hi}=100$~GHz 
for all components. We had to do the same for the calculation of the minimum energy magnetic 
field (column {\bf 11})  and also further assume a $\nu_{lo}$ for the non-nuclear components. 
This calculation also required, apart from the assumptions in the text, the major 
($\theta_x$; column {\bf 3}) and minor ($\theta_y$; column {\bf 4}) axis of each component 
as well as the guessed physical linear depth $s=\theta_y$ (column {\bf 5}). 
Finally, in column {\bf 12}, we also present the minimum energy density.}
\begin{center}
\begin{tabular}{ccccccccccccc} \hline
{\bf (1)} & {\bf (2)} & {\bf (3)} & {\bf (4)} & {\bf (5)} & {\bf (6)} & {\bf (7)} & {\bf (8)} & {\bf (9)} & {\bf (10)} & {\bf (11)} & {\bf (12)} & {\bf (13)} \\
Comp. &	Data	& $\theta_x$	&$\theta_y$&	$s$&	$\nu_{hi}$&	$\nu_{lo}$&	$\alpha_{\nu_{lo}}^{\nu_{hi}}$ $^{**}$&	$\nu_0$&	$S_{\nu_0}$&	$B_{me}$&	$u_{me}$&	$t_{sync}$	\\ 
	& &	(arcsec)	&(arcsec)	&(kpc)&		(GHz)&	(GHz)	& &	(GHz)	& (Jy)&	($\mu G$)&	(erg/cm$^3$)&	(yrs)  \\ \hline
N--small	&{\small VLBA 5 GHz}&	0.0008&	0.00026&	0.00081&		100	&5&	0.52$_{5}^{43}$&	4.983&	0.121	& $5\times 10^2$ & $2\times 10^{-8}$ &	$1\times 10^{4}$ \\
N--large	&{\small VLA-A 8.4 GHz}&	0.0496&	0.000496&	0.00154&		100	&5&	0.68$_{5}^{8.4}$&	8.452&	0.115	& $1\times 10^2$ 	& $1\times 10^{-9}$ 	& $1\times 10^{5}$ \\
A	&{\small VLA-A 43 GHz}	& 0.0838	&0.000838&	0.0026&		100	&0.1&	0.66$_{5}^{43}$&	43.34&	0.006&	$6\times 10^1$ 	& 	$3\times 10^{-10}$ 	& $3\times 10^{5}$ \\
C	&{\small MERLIN 5 GHz}&	0.0607	&0.0103&	0.032&		100	&0.01&	1.46$_{1.7}^{5}$&	4.994&	0.007	& $3\times 10^1$ & $1\times 10^{-10}$ 	& $7\times 10^{5}$ \\ \hline
\end{tabular}
\end{center}
\label{MagTime}
\end{table*}

\[
t_{sync}=1.610 \times 10^9 \frac{ B_{me}^{0.5} }{B_{me}^{2} + B_{CMB}^{2} } \frac{1}{ \nu_{hi}^{0.5} (1+z)^{0.5}} \:\: {\rm (yrs)}
\]
and we obtain, ignoring the CMB contribution and applying the redshift of B2151+174:
\[
t_{sync}\simeq \frac{ 1.452 \times 10^9 }{ \nu_{hi}^{0.5} B_{me}^{1.5} } \:\: {\rm (yrs)} \: .
\]
This results in timescales up to $\sim7\times10^{5}$~yrs, about two orders of magnitude higher than the crude expansion lower limit. This might be due to the presence of relevant expansion effects (e.g.\ \pcite{Muretal99}).

The radio timescales can also be estimated from the general properties of the larger
cluster samples. Taking A2390 as a unique ``out-bursting'' source in the sample of
Ball et al. (1993) then approximately one per cent of all cDs are in this state
at any one time. In the extreme limit of each system experiencing only one outburst in the 
lookback time range sampled (3$\times 10^9$ yrs) then the average duration of these outbursts
is $\sim3\times 10^7$ yrs (comparable to the duration of injections assumed by Dalla Vecchia et al. (2004) of $10^7$ yrs).
If they happen more often, then this duration will be lower.
Obtaining the statistics of a large and complete sample of central cluster radio sources
with multi-frequency radio mapping should allow much better limits on the timescales
of outbursts and their duty-cycles.

\subsection{The jet direction and galaxy structure}

The orientation of the jet components close to north-south is in stark contrast
to what is observed in HST images of the cD of Abell 2390 which show a biconical structure,
strongly suggestive of nuclear ionization, oriented at p.a.  --45$^{\circ}$, the same orientation
of a sharp linear dust lane that runs across the galaxy \cite{Edgetal99}.
One would expect the radio jet to lie in the same p.a., between the projected limits of 
the ionisation cone (e.g. \pcite{BreFabCra97}) but there is a 45$\degr$ misalignment.

The most obvious explanation for this misalignment is that the structure
seen in the HST imaging is related to a previous outburst when the jet was orientated
differently. The precession of the supermassive black hole and associated accretion
disk is possible. The displacement of component A from the near axisymmetry of 
the nucleus and components B and C (170$^{\circ}$ with respect to the nucleus) 
may be smaller scale evidence for such precession, as well as the twisting northern jet suggested by Figures~2a,b,c, but the effects of beaming
must be taken into account (e.g.\ from Figure~1a and Table~2, $S_A/S_C \sim 4$).

\subsection{The origin of steep spectrum emission}

\begin{figure} 
\setlength{\unitlength}{1cm}
\begin{picture}(8,9)
  \put(-1.8,9.5){\includegraphics{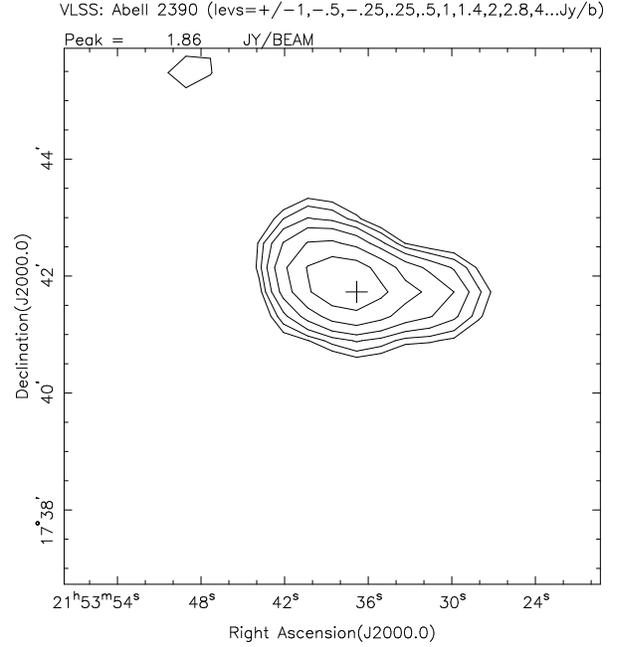}}
\end{picture}
\caption{Abell 2390 as seen with the VLSS  at 74~MHz ({\tt lwa.nrl.navy.mil/VLSS}).  The component is fitted with a 
$\sim1.8\arcmin$ size in the east-west direction but has a still unknown size in the north-south direction ($<0.7\arcmin$).} 
\label{VLSS}
\end{figure}

A recent VLA Low-frequency Sky-Survey (VLSS) 74~MHz map (Figure~\ref{VLSS}) shows a significant, extended source
(modeled as $\simeq1.8\arcmin \; \times<0.7\arcmin$ in size; at the 0.228 redshift of Abell 2390 \cite{LeBetal91,Yeeetal96}  
this translates into $\simeq  0.33 \;\times <0.12$~Mpc)
with a flux density that lies well above the extrapolation of the spectral index of the components resolved at higher frequencies (Figure~3).
The extent to the West matches both the X-ray morphology,
which shows an excess in this direction, and the lensing models, that show evidence for an additional mass
component. This low frequency emission is inconsistent with the low frequency contribution from the `N-Large'
components we resolve in this study (Figure~4) but is consistent with the diffuse component found in the 1.4~GHz
VLA map of Bacchi et al.\ (2003) that is too diffuse to be resolved in our observations.
Assuming that the majority of the lowest frequency data arises from this diffuse component then
literature data give $\alpha_{74}^{365}\simeq 1.10$ (Figure~3).
This diffuse, steep spectrum component implies the presence of a mini-halo like the one observed
in Perseus/3C84 (e.g.\ Silver et al. 1998; Pedlar et al. 1990) and A2052/3C317 (Zhao et al.\ 1993, Venturi et al.\ 2004)
and points to a history of activity in this system over a period much longer than the current event.

This result highlights the importance of low frequency studies of cluster cores in addition to the
more conventional higher frequency studies to gauge the power output and duty cycle of the radio sources
that are purported to provide the energy required to reduce or balance cooling.

\section{Conclusions}

The radio source associated with the cD in the rich, X-ray luminous cluster A2390
(B2151+174) is unusual in a number of important aspects. First, it is one
of the most compact and flat spectrum sources in a cluster core known.
Second, it shows a complex, compact twin-jet structure. Third, the orientation
of the jets is not aligned with apparent structure of the host galaxy on larger scales.

These properties are intriguing given the recent realisation that 
the influence of an AGN in the centre of a cluster could have
a very significant impact on the intracluster medium in the core
and the suppression of cooling. B2151+174 may be an
example of the early stage of a `bubble' being blown into the
ICM where the plasma is yet to expand. This initial phase
will be relatively short (10$^{3-4}$yrs) compared to the later
stages of the expansion (10$^{6-7}$yrs) so few `young' sources
can be expected in any selection of clusters. 

It is important to note that the 1.4~GHz radio power of B2151+174 ($10^{25.1}$~W/Hz) is
relatively small compared to the events that are thought to
have an appreciable effect on the ICM in the core of
a cluster \cite{dlVetal04} but the probability of
observing the early stages of a sufficiently powerful
event is small.  In samples of clusters of 
galaxies at least 300 are needed before
the meaningful statistics of these `young' sources can be assessed
as the ``outburst'' period is a small fraction of the duration of
each injection event.
Alternatively, establishing what fraction of bright, compact
Giga-Hertz peaked spectrum radio sources (the probable appearance
of such sources) are in cluster cores would also provide 
important constraints on these issues.

Perhaps the most puzzling aspect of B2151+174 is the 
gross misalignment of the smallest scale radio jet with 
the larger scale `ionization cones' and dust disk
seen in HST imaging. This may simply result from precession
of the central supermassive black hole between two distinct
episodes of activity or indicate a complex change in the 
angular momentum of gas accreted in the inner part of the 
disk in this galaxy. High spatial resolution radio imaging
of other systems in which a clear large scale alignment
is found will be important in addressing this issue.

\section*{Acknowledgments}

We acknowledge Peter N. Wilkinson for helpful discussions. 
We also acknowledge the comments from an anonymous referee.
The authors acknowledge support from the Funda\c{c}\~ao para a
Ci\^encia e a Tecnologia (FCT) under the ESO programme: PESO/P/PRO/15133/1999.
PA acknowledges support for this research by the European
Commission under contract ERBFMGECT 950012 and  also
the research grant by the FCT Praxis XXI
BPD~9985/96. ACE thanks the Royal Society for generous support.

This paper has made use of the NASA Extragalactic Database (NED). 
The National Radio Astronomy Observatory is a facility of the National
Science Foundation operated under cooperative agreement by Associated
Universities, Inc. 
  MERLIN is
operated as a National Facility by JBO, University of Manchester, on
behalf of the UK Particle Physics \& Astronomy Research Council.



\end{document}